\documentclass[journal]{IEEEtran}
\IEEEoverridecommandlockouts
\usepackage{cite}
\usepackage{float}
\usepackage{stfloats}
\usepackage{fancyhdr}
\usepackage{color}
\usepackage{amsmath}
\usepackage{graphicx} 
\graphicspath{{./figures/}}
\usepackage{algorithm}
\usepackage{algorithmic}
\usepackage{epsfig}
\usepackage{subfigure}
\usepackage{txfonts}
\usepackage{setspace}
\usepackage{bm}
\begin{document}
	\title{Orthogonal-Time-Frequency-Space Signal Design for  Integrated Data and Energy Transfer: Benefits from Doppler Offsets
	}
	
	\author{\IEEEauthorblockN{
			Jie~Hu, \emph{Senior~Member, IEEE},
			Ke~Xu, \emph{Student~Member, IEEE},
			Luping Xiang, \emph{Member, IEEE},
			and~Kun~Yang}, \emph{Fellow, IEEE}
			\vspace{-0.5 cm}\\

		\thanks{Jie Hu is with the Yangtze Delta Region Institute (Huzhou) and the School of Information and Communication Engineering, University of Electronic Science and Technology of China, Huzhou, 313001, China.
  
  Kun Yang is with School of Computer Science and Electronic Engineering (CSEE), University of Essex, Essex CO4 3SQ, U.K.

All the authors are also with the School of Information and Communication Engineering, University of Electronic Science and Technology of China, Chengdu 611731, China, email: hujie@uestc.edu.cn, 202121010634@std.uestc.edu.cn, luping.xiang@uestc.edu.cn, kyang@ieee. or g.  (\textit{Corresponding Author: Luping Xiang.})
		}
	}
	\maketitle
	


	\begin{abstract}
	Integrated data and energy transfer (IDET) is an advanced technology for enabling energy sustainability for massively deployed low-power electronic consumption components. However, the existing work of IDET using the orthogonal-frequency-division-multiplexing (OFDM) waveforms is designed for static scenarios, which would be severely affected by the destructive Doppler offset in high-mobility scenarios. Therefore, we proposed an IDET system based on orthogonal-time-frequency-space (OTFS) waveforms with the imperfect channel assumption, which is capable of counteracting the Doppler offset in high-mobility scenarios. At the transmitter, the OTFS-IDET system superimposes the random data signals and deterministic energy signals in the delay-Doppler (DD) domain with optimally designed amplitudes. The receiver optimally splits the received signal in the power domain for achieving the best IDET performance. After formulating a non-convex optimisation problem, it is transformed into a geometric programming (GP) problem through inequality relaxations to obtain the optimal solution. The simulation demonstrates that a higher amount of energy can be harvested when employing our proposed OTFS-IDET waveforms than the conventional OFDM-IDET ones in high mobility scenarios.
			\end{abstract}
	\section{Introduction}
\IEEEPARstart{I}{n} the upcoming era of Internet of Everything (IoE), it is impossible to recharge massively deployed miniature communication devices by wire. To maintain the energy sustainability of communication devices, a promising solution is composed by wireless data transfer (WDT) and wireless energy transfer (WET), where the rectifier at the receiver converts the radio frequency (RF) signal into the direct current (DC) to be stored in energy-buffers or to be invoked for directly driving electronic loads. By splitting the received signal in the time or power domain, energy harvesting (EH) and data receiving can be achieved simultaneously.

Over the past decades, numerous works have investigated the signal designs for enabling  integrated data and energy transfer (IDET)~\cite{hujiesurvey}.  For example, the joint power and subcarrier allocation  was considered in the IDET system with multiple users \cite{powerallocation}. The coding controlled \cite{hujiecoding}  and constellation rotation  aided  \cite{ConstellationRotation} IDET  were proposed to improve the EH performance while ensuring the data transmission rate. A practical IDET system based on the phase-shift modulation was studied in \cite{phase-shift}.
Additionally, the signal design for both  the non-linear and linear energy harvesters was  studied in \cite{208clerckx}.

As discussed above, the existing IDET signal designs were almost based on the classic orthogonal-frequency-division-multiplexing (OFDM). Moreover,
the OFDM-IDET performance was degraded   caused by the frequency offset in the high mobility scenario had proved in  \cite{OFDM_freoff}.
Therefore, we urgently need an IDET system to combat the adverse effect of the frequency offset induced by high mobility. Fortunately, orthogonal-time-frequency-space (OTFS)  modulation \cite{2017OTFS}, designed in the delay-Doppler (DD) domain rather than in the time-frequency (TF) domain, was proposed to address the frequency offset issue incurred by the Doppler effect.

However, to our best knowledge, research on OTFS-IDET is still a blank in the literature, which motivates our work in this paper. In summary,	our novel contributions are as follows:
	\begin{itemize}
		\item An OTFS-IDET  system is proposed. The transmitter sends superimposed random data signals and deterministic energy signals to formulate the optimal IDET waveforms, while the receiver splits the received signal and simultaneously  performs the EH and data recovery.
		\item Considering the imperfect channel estimation, we jointly optimise the amplitudes of the data and energy signals at the transmitter and the power splitting ratio at the receiver to maximise the harvested energy, while satisfying the data rate requirement. By adopting inequality relaxation, we then transform the original non-convex optimisation problem into a geometric programming (GP) problem.
		\item Simulation results demonstrate the performance gain of our proposed OTFS-IDET over the conventional OFDM-IDET in high-mobility scenarios. 
	\end{itemize}

    The rest of this paper is organized as follows: Section II introduces the system model, which is followed by the OTFS-IDET signal design in Section III. After providing simulation results in Section IV, our paper is concluded in Section V.

\begin{figure*}[!t]
		\centering
	\includegraphics[width=160mm]{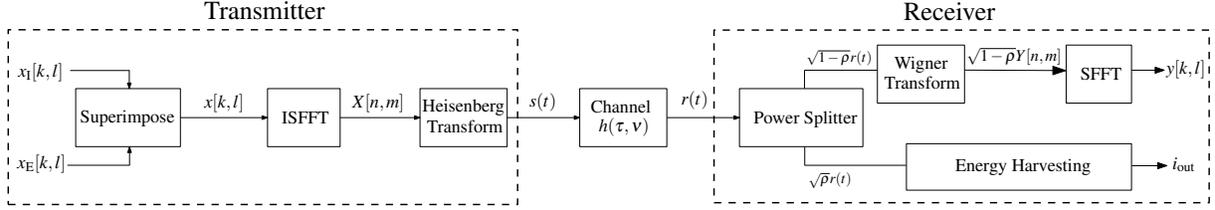}
		\caption{System transmitter and receiver.} \label{fig:system}
\end{figure*}

\section{System Model}
\subsection{OTFS-IDET Transmitter}
\label{sec.transmitter}

We study an OTFS-IDET system with a single pair of transmitter and receiver, as illustrated in Fig.~\ref{fig:system}. 
$x_\textrm{I}$ is the $K \times L$  data signal matrix with the elements $x_\textrm{I}[k,l]=z_\textrm{I}[k,l] x_\textrm{D}[k,l]$, where the $z_\textrm{I}  \sim \mathcal {CN}(0,1)$ is the random information matrix with the elements $z_\text{I}[k,l]$ and the $x_\textrm{D}$ is the matrix of power control with the element $x_\text{D}[k,l]$.  $x_\textrm{E}$ is a deterministic $K \times L$ energy signal matrix with the elements $x_\textrm{E}[k,l]$. At the transmitter, both are generated in the DD-domain, where $k=0, 1, \cdots, N-1$ and $l=0, 1, \cdots, M-1$ are the Doppler offset index and delay index, respectively. Here, $M$ and $N$ denote the number of the subcarriers and the number of the time slots, respectively. 
The superimposed DD-domain signal can be then expressed as:
	\begin{align}
			x[k, l]=&x_{\textrm{I}}[k, l]+x_{\textrm{E}}[k, l].
	\end{align}

Next, the ISFFT converts the DD-domain signal to the TF-domain, which can be expressed as:
	\begin{align}
			X{[n, m]}=&\frac{1}{\sqrt{N M}} \sum_{k=0}^{N-1} \sum_{l=0}^{M-1} x[k, l] e^{j 2 \pi\left(\frac{n k}{N}-\frac{m l}{M}\right)} \nonumber \\
			=&\frac{1}{\sqrt{N M}} \sum_{k=0}^{N-1} \sum_{l=0}^{M-1} (x_\textrm{I}[k, l]+x_\textrm{E}[k, l] )e^{j 2 \pi\left(\frac{n k}{N}-\frac{m l}{M}\right)}\nonumber \\
			=&X_\textrm{I}{[n, m]}+X_\textrm{E}{[n, m]},\label{ISSFT}
	\end{align}
for $n=0, 1, \cdots, N-1$ and $m=0, 1, \cdots, M-1$. $X_\textrm{I}$ is the data signal matrix with the elements $X_\textrm{I}[n,m]$ and $X_\textrm{E}$ is the energy signal matrix with the elements $X_\textrm{E}[n,m]$ in the TF-domain. The transmit signal $s(t)$ can be obtained through the Heisenberg Transform. 


\subsection { Channel Model}
The channel\footnote{By adopting some beamformers and combiners, we can readily convert a multiple input multiple output (MIMO) channel to an equivalent  single input single output (SISO) channel.  The OTFS signal can be then designed in the DD-domain by using our method.} fading coefficient $h(\tau,\nu)$ with the imperfect channel estimation in DD-domain is modelled as \cite{Impefectchannel}:
	\begin{align}
				h(\tau, \nu)=\sum_{i=0}^{P-1} \left(\hat h_{i}+e_{i}  \right) \delta\left(\tau-\tau_{i}\right) 	\delta\left(\nu-\nu_{i}\right),
	\label{eq.channel1}
	\end{align}
where $P$ denotes the number of propagation paths, and $\hat h_{i}$
represents the estimated channel coefficient on the $i$-th path with an estimation error of $e_i\sim \mathcal{CN}(0,MN\sigma_e^2/P)$. Moreover, $\tau_{i}$ and $\nu_{i}$ represent delay and Doppler offset of the $i$-th path, respectively. The parameters $\tau=\frac{1}{M \Delta f}$ and $\nu=\frac{1}{N T}$ represent the resolutions of the DD-domain. Note that $\{M,\Delta f,N,T\}$  are set in the TF-domain, which are the same as the OFDM counterpart. Specifically, $T$ and $\Delta f$ represent the symbol duration and the subcarrier frequency interval, respectively. Accordingly, the signal bandwidth is $M\Delta f$, while the total time duration is $NT$.


The corresponding channel coefficient in the TF-domain can be then expressed as:
	\begin{align}
			H[n, m]=\int_{\tau} \int_{\nu} 	h(\tau, \nu) e^{j 2 \pi(\nu nT-m \Delta f \tau)} d \tau d \nu.
	\end{align}
By considering the channel  estimation error in the DD-domain, $ H[n,m]$ can be further expressed as:
\begin{align}
         H[n,m]=\hat H[n,m]+ e[n,m],
\end{align}
where  estimated channel coefficient in the TF-domain is $\hat H[n,m]$ with the channel estimation error  $e[n,m]\sim \mathcal {CN}(0,\sigma_e^2)$.

\subsection{OTFS-IDET Receiver}
\label{sec.receiver}

The ideal pulse shape\footnote{The practical pulse shape of OTFS may introduce extra inter-carrier and inter-symbol interference in our model \cite{practical_shape}, which would be considered in our future work.} is considered in this paper.  By invoking the Wigner transform on the received signal $r(t)$, the channel input-output relation in the TF-domain can be expressed as: 
\begin{align}
\begin{split}
	&{Y}[n,m]= X[n,m]H[n,m]+N_{\text{0}}[n,m],\\
	&= \left(X_{\text{I}}[n,m]+X_{\text{E}}[n,m]\right)\left(\hat H[n,m]+e[n,m]\right)+N_{\text{0}}[n,m],\\
	&= Y_{\text{I}}[n,m]+ Y_{\text{E}}[n,m]+N_{\text{0}}[n,m],
\end{split}
\end{align}
where $N_{\text{0}}[n,m]$ is the  additive white Gaussian noise (AWGN), $Y_{\text{I}}[n,m]=X_{\text{I}}[n,m]\left(\hat H[n,m]+e[n,m]\right)=\hat Y_{\text{I}}[n,m]+X_{\text{I}}[n,m]e[n,m]$ and $Y_{\text{E}}[n,m]=X_{\text{E}}[n,m]\left(\hat H[n,m]+e[n,m]\right)=\hat Y_{\text{E}}[n,m]+X_{\text{E}}[n,m]e[n,m]$.

By denoting the power splitting ratio as $\rho\in\left[0,1\right]$ at the receiver, we have a portion of received signal $\sqrt{\rho}{Y}[n,m]$ for EH, while the remaining  $\sqrt{1-\rho} {Y}[n,m]$ for data recovery. 
After performing SFFT, the received signal $ y[k, l]$ in the DD-domain is expressed as:
	\begin{align}
		\begin{split}
			 y[k, l]=\frac{1}{\sqrt{N M}} \sum_{n=0}^{N-1} \sum_{m=0}^{M-1}\sqrt{1-\rho} {Y}[n, m] e^{-j 2 \pi\left(\frac{n k}{N}-\frac{m l}{M}\right)}.
		\end{split}
	\end{align}

\section{OTFS-IDET Signal Design}
\subsection{WET Performance Analysis}
\label{sec.wet}
	In this work, we adopt a diode based nonlinear\footnote{The linear EH model cannot accurately capture the relationship between the output DC $i_{\text{out}}$ and
the input signal $Y$, we decide to design our OTFS signal by considering the non-linear EH model of
Eq.~\eqref{NLinear}.} EH model energy harvester \cite{208clerckx}, which leverages the output DC to evaluate the WET performance.  
	We ignore the contribution of the noise to the WET performance, since it is far lower than the signal power.
For clarity, $Y$, $Y_{\text{I}}$, $ Y_{\text{E}}$, $H$  and $\hat H $ are defined as the matrices with the elements   $Y[n,m]$, $Y_{\text{I}}[n,m]$, $Y_{\text{E}}[n,m]$, $ H[n,m]$ and $\hat H[n,m]$, respectively.
 
First, the random information matrix $Z_\textrm{I}$ in the  TF-domain  is obtained through $z_\textrm{I}$ by Eq.~\eqref{ISSFT}. Second, the corresponding power control matrix for $Z_\textrm{I}$ in the TF-domain is defined as $X_\textrm{D}$, which satisfies:    $Z_\textrm{I}\otimes X_\textrm{D}=\textrm{ISFFT}\left(z_\textrm{I}\otimes x_\textrm{D}\right )=X_{\textrm{I}}$. Third, the received signal for WDT is expressed as $Y_{\text{I}}=X_{\text{I}}\otimes H=Z_\textrm{I}\otimes X_\textrm{D}\otimes H$. Finally, the receive power control matrix for WDT is obtained as $Y_{\text{D}}=X_{\text{D}}\otimes H$. By considering the channel estimation error, we have $\hat Y_{\text{D}}=X_{\text{D}}\otimes \hat H$.

 The relationship between the output DC $i_{\text{out}}$ and the input RF signal $Y$ is then expressed as: 
	\begin{align}
		i_{\text{out}}=k_{2}R\psi\left(\left(\sqrt{\rho}Y\right)^{2}\right)+k_{4}R^{2}\psi \left(\left(\sqrt{\rho}Y\right)^{4}\right),
		\label{eq.current1}
	\end{align}
where $k_{2}$, $k_{4}$ and $R$ are the circuit parameters, $\psi(\cdot)$ is the DC component of the received signal. Since random and deterministic signals behave differently in nonlinear EH model, Eq.~\eqref{eq.current1} can be further reformulated as: 
	\begin{align}
		\begin{split}
			i_{\text {out }}&=k_{2} \rho R \left(\psi \left(Y_{\text{E}}^{2}\right)+\psi \left(Y_{\text{D}}^{2}\right)\right)\\&+k_{4} \rho^{2} R ^{2}\left(\psi \left(Y_{\text{E}}^{4}\right)+\psi \left(Y_{\text{D}}^{4}\right)+6 \psi \left(Y_{\text{E}}^{2}\right) \times \psi \left(Y_{\text{D}}^{2}\right)\right),
		\end{split}\label{NLinear}
	\end{align}
where we have:
\begin{subequations}
	\begin{align}
		\psi \left(Y_{\text{D}}^{2}\right)=&\frac{1}{2} \sum_{n=0}^{N-1} \sum_{m=0}^{M-1}\left(\left| \hat Y_{\text{D}}[n, m]\right|^{2}+ |X_{\textrm{D}}[n,m]|^2 \sigma_e^2\right)        ,
		\\
		\psi \left(Y_{\text{E}}^{2}\right)=&\frac{1}{2}\sum_{n=0}^{N-1} \sum_{m=0}^{M-1}\left(\left|\hat Y_{\text{E}}[n, m]\right|^{2}+|X_{\textrm{E}}[n,m]|^2 \sigma_e^2\right) ,
		\\
		\begin{split}
		\psi \left(Y_{\text{E}}^{4}\right)=&\frac{3}{8}\sum_{{n}=0}^{{N-1}} \sum_{\substack{m_{0}, m_{1}, m_{2}, m_{3} \\ m_{0}+m_{1}=m_{2}+m_{3}}}^{{M-1}}\left(\prod_{j=0}^{3}  \left| \hat Y_{\text{E}}[n, m_j]\right|+\prod_{j=0}^{3} |X_{\textrm{D}}[n,m_{j}]| \sigma_e \right)
		\end{split}.
			\\
        \begin{split}
		\psi \left(Y_{\text{D}}^{4}\right)=&\frac{3}{4}\sum_{n=0}^{N-1} \left(\sum_{m=0}^{{M-1}}\left|\hat Y_{\text{D}}[n, m]\right|^{2}+|X_{\textrm{D}}[n,m]|^2 \sigma_e^2\right)^2.
		\end{split}
	\end{align}\label{DC2}
\end{subequations}

	\subsection{WDT Performance Analysis}
	\label{sec.wit}
The WDT performance of the OTFS-IDET system is characterized in terms of the data rate. At the transmitter, randomly modulated signals in the DD-domain is converted to the TF-domain via the ISFFT, which still obey Gaussian distribution. 

The received signal is split in the power domain at the receiver, the remaining signal power $P_{\text{D}}{[n,m]}$ for data recovery is expressed as:
	\begin{align}
		P_{\text{D}}{[n,m]}=(1-\rho) \left(|\hat Y_{\text{D}}{[n,m]}|^{2}+|X_{\rm{D}}[n,m]|^{2}\sigma_e^2\right),
	\end{align}
where 	$\hat P_{\text{D}}{[n,m]}=(1-\rho)|\hat Y_{\text{D}}{[n,m]}|^{2}$ is the useful signal power, and $(1-\rho)X_{\rm{D}}[n,m]\sigma_e^2$ is the interference caused by the channel estimation error.
 Similarly, the power $P_{\text{E}}{[n,m]}$ of the received energy signal that may interfere the data recovery is expressed as: 

	\begin{align}
			P_{\text{E}}{[n,m]}=(1-\rho) \left(|\hat Y_{\text{E}}{[n,m]}|^{2}+|X_{\rm{E}}[n,m]|^2\sigma_e^2\right).
	\end{align}
The parameter $\lambda$ is defined as the remaining interference  caused by the energy   single on the data recovery at the receiver. Hence, the interference at the  $n$-th time slot and the $m$-th subcarrier can be expressed as $\lambda P_{\text{E}}[n,m]$.
As a result, the data rate $C_{n}$  at the $n$-th time slot can be characterized by the classic Shannon-Hartley capacity, which is expressed as:

\begin{align}
 \begin{split}
		C_{n}=\sum_{m=0}^{M-1} \log_{2} \Bigg(1+&\Big(\frac{\lambda (1-\rho) \left(|\hat Y_{\text{E}}{[n,m]}|^{2}+|X_{\rm{E}}[n,m]|^2\sigma_e^2\right)}{(1-\rho)|\hat Y_{\text{D}}{[n,m]}|^{2}}\\+&\frac{(1-\rho)|X_{\rm{D}}[n,m]|^2\sigma_e^2+P_{\rm{noise}}}{(1-\rho)|\hat Y_{\text{D}}{[n,m]}|^{2}}\Big)^{(-1)}\Bigg)
  \end{split}
\end{align}

 where 	$P_{\rm{noise}}$ represents the Gaussian noise power. Hence, the average data rate $C$ per time-slot can be expressed as: 
    \begin{align}
    	C=\frac{\sum_{n=0}^{N-1} C_{n}}{N}.
    \end{align}

\subsection{Optimal Signal Design}

Given a budget of total transmit power $P_{\rm{tx}}$, the amplitudes of transmit data signal $x_{\text{D}}$ and  the energy signal $x_{\text{E}}$  in the DD-domain at the transmitter and the power splitting ratio $\rho$ at the receiver should be jointly designed for maximising the output DC $i_{\rm{out}}$, while satisfying the  minimum data  rate requirement $R_{\rm{min}}$. The fixed power of the pilot used for the channel estimation is assumed as $P_{\rm{p}}$.  Hence, the corresponding optimisation problem is formulated as:
	\begin{align}
		(P1): \mathop {\arg \max }\limits_{{x_\text{D}}, {x_\text{E}}, \rho }& \quad {i_{\text{out}}}({x_\text{D}},{x_\text{E}},\rho ) \label{X}\\
		\rm{s.t} \quad & {C}({x_\text{D}},{x_\text{E}},\rho ) \ge {R_{\mathrm{min}}},\tag{\ref{X}{a}} \\
		&\frac{1}{2}\left[ {||{x_\text{D}}||_{m_2}^2 + ||{x_\text{E}}||_{m_2}^2} \right] \le {P_{\rm{o}}},\tag{\ref{X}{b}} \\&
	    \sum_{m=0}^{M-1} f_{\text{1}}[n,m] \le P_{\rm{peak}}, \forall n , \tag{\ref{X}{c}}\\&
		\rho  \in [0,1].\tag{\ref{X}{d}}
	\end{align}
where $P_{\rm{o}}=P_{\rm{tx}}-P_{\rm{p}}$ is the power limitation for the data and energy transfer, and $P_{\rm{peak}}$ is the maximum peak power in each time slot that can be achieved by the power amplifier at the transmitter. The  $f_{\text{1}}[n,m]=|X_{\text{D}}[n,m]|^{2}+ |X_{\text{E}}[n,m]|^{2}$, and $m_2$ represents the Frobenius norm.
	
Unfortunately, this is a non-convex optimisation problem. Then, we transform (P1) to a GP \cite{GP} to obtain a solution with a reasonable computational complexity. Note that both the WDT and WET performance can be expressed in the TF-domain. And, the optimisation variables $x_{\text{D}}{[k,l]}$ and $x_{\text{E}}{[k,l]}$ can be derived by  the  $X_{\text{D}}{[n,m]}$ and  $X_{\text{E}}{[n,m]}$, respectively.

Hence, the objective of (P1) becomes ${i_{\rm{out}}}({{X}_\text{D}},{{X}_\text{E}},\bf{\rho} )$. Let $g_{k_1}({{X}}_\text{D}, {{X}}_\text{E},\rho)$ represents the monomial terms of it, where ${i_{\rm{out}}}({{X}_\text{D}},{{X}_\text{E}},\bf{\rho})$ $=\sum_{k_1=0}^{K_1-1} g_{k_1}({{X}}_\text{D}, {{X}}_\text{E},\rho)$. In the same way, the $g_{k_2}({{X}}_\text{D}, {{X}}_\text{E},\rho)$ is defined as the momomial of $\kappa{[n,m]}$ satisfying  $\kappa 
{[n,m]}=\sum_{k_2=0}^{K_2-1} g_{k_2}({{X}}_\text{D}, {{X}}_\text{E},\rho)=\lambda P_{\text{E}}[n,m]+P_{\text{D}}[n,m]+P_{\textrm{noise}}$.
The $K_1 $ and $K_2$ respectively represent that the number of monomials in  ${i_{\rm{out}}}({{X}_\text{D}},{{X}_\text{E}},\bf{\rho} )$ and  $\kappa 
{[n,m]}$. 

Then the data rate constraint (\ref{X}a) can be expressed as: $2^{R_{\text{min}}} \prod\limits_{m = 0}^{M - 1}\left(\lambda P_{\text{E}}[n,m]+(1-\rho)X_{\rm{D}}[n,m]\sigma_e^2+P_{\rm{noise}}\right)/{\kappa [n,m]} \leq 1, \forall n$. Furthermore, by introducing the parameter $\gamma [m]$ and $f_{\rm{2}}[n,m]=\left(\lambda P_{\text{E}}[n,m]+(1-\rho)X_{\rm{D}}[n,m]\sigma_e^2+P_{\rm{noise}}\right) / \gamma  [m]$, the constraint can be divided it into two parts: $2^{R_{\rm{min}}} \prod \limits_{m = 0}^{M - 1} f_{\rm{2}}[n,m] \leq 1$, $\forall n$ and  $\gamma [m]/\kappa [n,m] \leq 1$, $\forall n$, $\forall m$.


With the variable $\eta$, the original maximisation problem can be transformed into a minimisation of $1/\eta$ with the  constraint ${\eta / {{{{\rm{ { i_{\rm{out}}}({{X}_\text{D}},{{X}_\text{E}},\bf{\rho} )}}}}}} \le 1$ .
In order to satisfy the form of GP, the summation in the denominator is converted to the multiplication via the  arithmetic mean-geometric mean (AM-GM) inequality as:
\begin{equation}
\begin{aligned}
	&{\hat i_{\rm{out}}}({{X}_\text{D}},{{X}_\text{E}},\rho )=\prod_{k_1=0}^{K_1-1}\left(\frac{g_{k_1}({{{ X}}_\text{D}}, {{{ X}}_\text{E}},\rho)}{t_{k_1}}\right)^{-t_{k_1}}\le {i_{\rm{out}}}({{X}_\text{D}},{{X}_\text{E}},\rho ) ,\label{hatI}
\end{aligned}	
\end{equation}
\begin{equation}
\begin{aligned}
	&\hat \kappa[n,m] =\prod_{k_2=0}^{K_2-1}\left(\frac{g_{k_2}({{{ X}}_\text{D}}, {{{ X}}_\text{E}},\rho)}{w_{k_2}}\right)^{-w_{k_2}}\le \kappa {[n,m]},\label{hatK}
\end{aligned}
\end{equation}
where $t_{k_1}$, $w_{k_2}$ satisfying $\sum t_{k_1}=1$, $\sum w_{k_2}=1$ are the weights of the above two inequalities. Now, the standard GP can be formulated as:
		\begin{align}
				(P2):\mathop {\arg\min}\limits_{{{X}_\text{D}},{{X}_\text{E}},\rho ,\bar \rho} &\quad\frac{1}{\eta}  \label{Z}\\
				\label{eq.problem2a}
		\rm{s.t} \quad &{\eta /{{{{\rm{ {\hat i_{\rm{out}}}({{X}_\text{D}},{{X}_\text{E}},\bf{\rho} )}}}}}} \le 1,  \tag{\ref{Z}{a}}\\
						\label{eq.problem2b}		
				&2^{R_{min}} \prod\limits_{m = 0}^{M - 1} f_{\rm{2}}[n,m] \leq 1, \forall n ,\tag{\ref{Z}{b}}\\	
				\label{eq.problem2c}
		&\gamma [m]/\hat \kappa [n,m] \leq  1, \forall n ,\forall m , \tag{\ref{Z}{c}}\\
				\label{eq.problem2d}
		&\frac{1}{2}\left[ {||{{{ X}}_\text{D}}||_{m_2}^2 + ||{{{ X}}_\text{E}}||_{m_2}^2} \right] \le {P_{\rm{o}}},\tag{\ref{Z}{d}}\\
		&\sum_{m=0}^{M-1}  f_{\text{1}}[n,m]\le P_{\rm{peak}},\forall n ,\tag{\ref{Z}{e}}\\
		&\rho  + \bar \rho  \le {\rm{1}}. \tag{\ref{Z}{f}}
	\end{align}
where $ \bar{ \rho}$ represents the splitting ratio for data recovery.

	\begin{algorithm}[!t]
		\renewcommand{\algorithmicrequire}{\textbf{Input:}} 
		\renewcommand{\algorithmicensure}{\textbf{Output:}}
		\caption{Signal Design for OTFS-IDET}
		\footnotesize
		\begin{algorithmic}[1]
			\REQUIRE ~~\\
				The DD-domain channel estimation $\hat h(\tau, \nu)$; Minimum data rate $R_{\rm{min}}$;Total transmission power $P_{\rm{tx}}$; Pilot power $P_{\rm{p}}$;
				Maximum peak power $P_{\rm{peak}}$;
				Required accuracy $\epsilon$;
			\ENSURE ~~\\
			The OTFS data signal and energy signal amplitude $x_{\text{D}}{[k,l]}$ and $x_{\text{E}}{[k,l]}$;\
			The power splitting ratio $\rho$;
			\STATE Calculate the $\hat H[n,m]$ through $\hat h(\tau, \nu)$ ; Calculate the power for optimisation $P_{\text{o}}=P_{\text{tx}}-P_{\text{p}}$; the average fading of channel: $h_{\text{L}}=\sum_{n=1}^{N}\sum_{m=1}^{M}(\hat H[n,m]^2+\sigma^2)/{MN}$; the estimation of maximum transmission: $C_{\rm{max}}= \log \left(1+	P_{\text{o}}h_{\text{L}}/{P_{\rm{noise}}}\right)$; the initialise factor: $\zeta  =R_{\text{min}}/{C_{\rm{max}}}$.	
					
			\STATE Initialise ${{X}}_{\text{D}}{[n,m]}=\sqrt{P_{\text{o}}\zeta /{MN}}$, ${{X}}_{\text{E}}{[n,m]}=\sqrt{P_{\text{o}}\left(1-\zeta\right)/{MN}}$, $\rho=1-\zeta$, $\bar{\rho}=\zeta$, $i_{\rm{out}}^{(0)}=0$, $i=0$, $t_{k_1}={1}/{K_{1}}$, $w_{k_2}={1}/{K_{2}}$;
			\REPEAT
			\STATE Calculate $\hat{i}_{\rm{out}}$ and $\hat \kappa{[n,m]}$ by the formula Eq.~(\ref{hatI}) and Eq.~(\ref{hatK});
			\STATE Update ${{ X}}_\text{D}, {{ X}}_\text{E}, \rho, \bar{\rho}$ by solving (P2); 
			\STATE Calculate $i_{\rm{out}}^{(i+1)}$ by the formula Eq.~(\ref{eq.current1});
			\STATE Update $t_{k_1} $,$w_{k_2}$ by the formula Eq.~(\ref{ep.update});
			\STATE Update $i \leftarrow i+1$
			\UNTIL $|i_{\rm{out}}^{(i)}-i_{\rm{out}}^{(i-1)}| < \epsilon $
			\STATE $x_{\text{D}}{[k,l]}=\text{SFFT}\left(X_{\text{D}}{\left[n,m\right]}\right)$ and $x_{\text{E}}{[k,l]}=\text{SFFT}\left(X_{\text{E}}{[n,m]}\right)$
		\end{algorithmic}

	\end{algorithm}
 
The optimisation algorithm is summarized in Algorithm 1, in which the gap between the intermediate variables $\left\{ \hat{i}_{\rm{out}}, \hat \kappa{[n,m]} \right\} $ and the upper bound $\left\{{i}_{\rm{out}}, \kappa{[n,m]} \right\}$ will be narrow by constantly modifying the weight $t_{k_1}$ and $w_{k_2}$ to approach the optimal value. The updated formulas of $t_{k_1}$ and $w_{k_2}$ are given as:
\begin{align}
	 t_{k_1}=&\frac{g_{k_1}({{  X}}_\text{D},{{  X}}_\text{E},\rho)}{{i_{\rm{out}}}({{X}_\text{D}},{{X}_\text{E}},\bf{\rho} )},&
	 w_{k_2}=&\frac{g_{k_2}({{  X}}_\text{D},{  X}_\text{E},\rho)}{\kappa{[n,m]}}.\label{ep.update}
\end{align}

By invoking the Ellipsoid method \cite{Ellipsoid},   (P2) can be solved with the complexity $\mathcal{O}\left( \left(2MN+1\right)^2\right)$. Suppose that  $\delta$  iterations are required to let Algorithm 1 converge, the total complexity becomes $\mathcal{O}\left(\delta \left(2MN+1\right)^2\right)$.
	
	
\section{Numerical Results}
	The OFDM-IDET system is considered as the benchmark of our OTFS-IDET  system. The parameters of the nonlinear energy harvester are set as $k_2 = 0.0034$, $k_4 = 0.3859$, and $R = 50$ $\Omega$ \cite{208clerckx}. Additionally, the transmit power $P_{\text{tx}}$ and  $P_{\text{p}}$ are set as $36.1~\text{dBm}$ and $20~\text{dBm}$;  the receive antenna gain is 2 dBi; the large-scale path loss is -50 dB. The system parameters employed in our simulations are summarised in TABLE \ref{table1} as \cite{xianglaos}. Especially, the maximum delay index is set as $l_{\tau_\text{m}}=\tau_\text{m}M\Delta f=6$ with the maximum delay $\tau_\text{m}=0.3~\text{ms}$. The maximum Doppler index is set as $k_{\nu_\text{m}}=(f_{\text{c}}V_{\text{m}}N)/(V_{\text{c}}\Delta f)=6$, where  $V_{\text{m}}=300~\text{Kmph}$ is the maximum speed in the scenarios and $V_{\text{c}}$ is speed of the light.  Additionally, the Doppler tap of the $i$-th  path is generated with equal probabilities from the set $\{-k_{\nu_\text{m}},-k_{\nu_\text{m}}+1,\dots,0,1,\dots,k_{\nu_\text{m}}\}$, while the delay tap set is $\{1,\dots, l_{\tau_\text{m}}\}$ and $\tau_{0}=0$.
	\begin{table}[H]
       \vspace{-12pt}
		\centering
		\renewcommand{\arraystretch}{1.3} 
		\caption{Simulation Parameters}
		\label{table1} 
	\begin{tabular}{c|c}
		\hline
		Parameter                & Value                 \\ \hline\hline
		Carrier frequency $f_{\rm{c}}$ (GHz)   & 27          \\ \hline
		Subcarrier frequency interval $\Delta f$ (KHz) & 15  \\\hline
		OTFS frame size $[N,M]$         & [12,12] \\ \hline
		Number of paths  $P$        & 3                     \\ \hline
		UE speed (Kmph)           & 30,150,300            \\ \hline
		Channel        & Ideal \& Imperfect                 \\ \hline
	\end{tabular}
    \end{table}
\vspace{-12pt}
	\begin{figure}[H]
		\centering
		\includegraphics[width = 85mm]{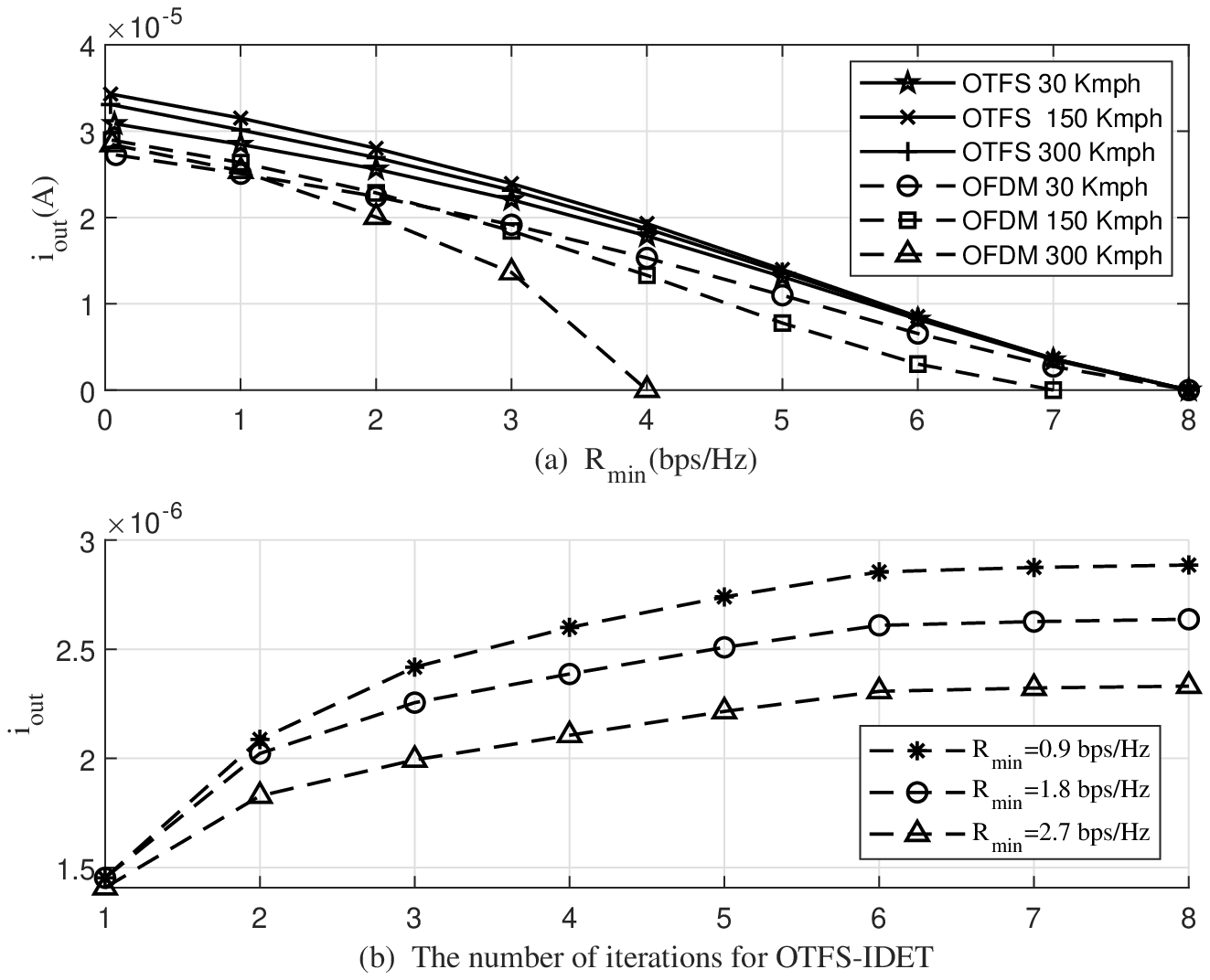}
\caption{(a) The performance comparison between OTFS-IDET and OFDM-IDET with different speeds, where $\sigma^2=0$; (b) The EH performance of the  OTFS-IDET system under the imperfect channel estimation.
}
\label{fig:OFFcompare}
	\end{figure} 

 First, the impact of the mobility on the OTFS-IDET and OFDM-IDET system is shown in Fig.~\ref{fig:OFFcompare}(a). Since, the Doppler offset cuasey speed makes the TF channel $H[n,m]$ fluctuate, the appropriate speed is beneficial to the OTFS-IDET system. 
 Doppler offset  leads to serious inter-carrier-interference (ICI) in the OFDM-IDET system. The OTFS signal can be designed by considering the channel characteristics in the DD-domain, while the OFDM signal can only be  designed in the frequency domain.  Therefore, the OTFS signal design with additional degree of freedom is expected to have a better IDET performance than the OFDM counterpart. Therefore, the OTFS-IDET system outperforms the OFDM-IDET counterpart in different speed scenarios.
Observe form Fig.~\ref{fig:OFFcompare}(b)  that the output DC $i_{\rm{out}}$ gradually increases as we have more iterations. Normally,  Algorithm 1 converges after 7 iterations.

	\begin{figure}[!t]
		\centering
		\includegraphics[width = 85mm]{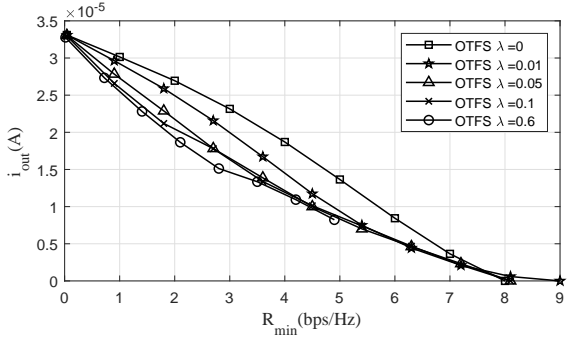}
    \caption{The performance comparison between OTFS-IDET and OFDM-IDET with different $\lambda$.  } 
    \label{fig:OTFS_lamuda}
    \vspace{-12pt}	
	\end{figure}
	
Second, we investigate the impact of the residual energy signal interference on the OTFS-IDET system in Fig.~\ref{fig:OTFS_lamuda}. Since, a larger value of $\lambda$ indicates more residual interference imposed by the energy signals on the data recovery at the receiver, the IDET performance is degraded, as shown in Fig.~\ref{fig:OTFS_lamuda}.
When the maximum data transmission rate is attained, the transmitter only sends the data signals without energy signal; in order to obtain the maximum DC, only the energy signals are transmitted, which results in varying $\lambda$ makes no difference in the output DC  performance with no requirement of data rate. Hence, all the curves share the same start and end points in the coordinates.
	\begin{figure}[!t]
		\centering
		\includegraphics[width =85mm]{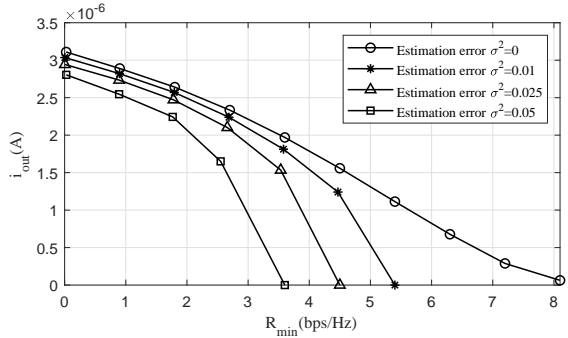}
    \caption{The EH performance of the OTFS-IDET sytem under the channel estimation error.  } 
    \label{fig:OTFS_IM}
    \vspace{-12pt}	
	\end{figure}

Finally, the OTFS-IDET performance with the imperfect channel estimation is portrayed in Fig.~\ref{fig:OTFS_IM}.  Observe from Fig.~\ref{fig:OTFS_IM} the performance is reduced, as the variance of the channel estimation error increases from $\sigma^2 = 0$ to $\sigma^2 = 0.05$. This is because the OTFS signal designed with the inaccurate channel estimation sacrifices the optimality to some extent.  Due to the estimation error, the performance of the OTFS-IDET system is reduced in terms of EH and data transmission.  When the gap between the  estimation channel and the perfect channel becomes larger, the optimised parameters become more unreasonable. Therefore, the performance of OTFS-IDET with the estimation error $\sigma^2=0.05$ is worse than the counterpart with $\sigma^2=0.01$.

	\section{Conclusion}
	We firstly studied an OTFS-IDET system and the corresponding data and energy signal design for maximising the harvested energy, by jointly optimising the signal amplitudes at the transmitter, as well as the power splitting ratio at the receiver, while meeting the data rate requirement of the receiver. The channel estimation error is considered in the system model. The original non-convex optimisation problem is transformed to a GP problem. The simulation shows that  the OTFS-IDET system performs better than the conventional OFDM-IDET counterpart, while the performance gain increases when the moving speed increases.

	\bibliography{refer} 
\end{document}